\documentclass[aps,prd,reprint,twocolumn,superscriptaddress,nofootinbib, showkeys]{revtex4-2}
\pdfoutput=1
\usepackage[utf8]{inputenc}
\usepackage[english]{babel}
\usepackage{amsmath, amssymb}
\usepackage{graphicx}
\usepackage{hyperref}
\usepackage{booktabs}
\usepackage{xcolor}

\usepackage{soul}

\begin{document}

\title{Sensitivity of polaron-molecule observables to MDR/GUP-like ultraviolet deformations at low energies via quantum computing}

\author{Ezequiel Valero}
\email[Author to whom any correspondence should be addressed: ]{ezequiel.valero@universidadeuropea.es}
\affiliation{Departament de Física, Universitat de Valencia, Valencia, Spain}
\affiliation{Escuela de Ciencias, Ingeniería y Diseño, Universidad Europea de Valencia, Paseo de la Alameda 7, 46010, Valencia, Spain}

\author{Hugo Catala}
\affiliation{Escuela de Ciencias, Ingeniería y Diseño, Universidad Europea de Valencia, Paseo de la Alameda 7, 46010, Valencia, Spain}
\affiliation{Instituto de Física Corpuscular, Universitat de València – Consejo Superior de Investigaciones Científicas, Parc Científic, E-46980 Paterna, Valencia, Spain}

\author{Victor Ilisie}
\affiliation{Escuela de Ciencias, Ingeniería y Diseño, Universidad Europea de Valencia, Paseo de la Alameda 7, 46010, Valencia, Spain}

\author{Germán Rodrigo}
\affiliation{Instituto de Física Corpuscular, Universitat de València – Consejo Superior de Investigaciones Científicas, Parc Científic, E-46980 Paterna, Valencia, Spain}
\begin{abstract}
We show that impurity many-body observables can display enhanced sensitivity to ultraviolet deformations of generalized-uncertainty-principle and modified-dispersion-relation type at accessible energy scales. Using a deformed polaron–molecule Hamiltonian constructed to preserve the infrared sector, we quantify the impact of such deformations on spectral and Ramsey observables and implement the corresponding dynamics in a controlled quantum-computing setting. We identify regimes near the polaron–molecule crossover where small ultraviolet deformations are strongly amplified, leading to experimentally resolvable changes in quasiparticle properties and spectral response. Our results establish a concrete sensitivity-based route to low-energy quantum-gravity phenomenology in a well-defined many-body platform and delimit the validity of the effective description. Furthermore, we report the experimental validation on the QRed superconducting quantum processor (BSC-CNS).
\end{abstract}


\maketitle
\section{Introduction}
A common expectation across candidate approaches to quantum gravity is that the standard continuum
description of spacetime may fail at sufficiently short distances, potentially inducing departures
from conventional quantum mechanics and relativistic kinematics
\cite{KonishiPaffutiProvero1990,Maggiore1993a,Maggiore1993b,KempfManganoMann1995}.
Over the last decades, this possibility has been explored within the broader framework of
quantum-gravity phenomenology, where potential Planck-scale effects are parametrized effectively and
confronted with observations and experiments
\cite{AmelinoCameliaEtAl1998}.

Among the  widely studied phenomenological parametrizations are modified dispersion relations
(MDR) and generalized uncertainty principles (GUP), which take the  possible ultraviolet (UV)
departures from standard kinematics through higher-order momentum corrections, deformed commutator
structures, or effective minimal-length scales
\cite{ScardigliCasadio2003,
HusainKothawalaSeahra2013,BossoLuciano2021,Bosso2024}. The main obstacle for direct tests is that
such effects are generically suppressed by the Planck scale. As a result, progress relies not only
on precision bounds, but also on identifying physical systems and observables that are especially
sensitive to controlled UV deformations
\cite{LiberatiVisser2013,Hossenfelder2013,AmelinoCamelia2013}.

A useful formulation of this strategy was given by Das and Vagenas, who showed that broad classes of
quantum-gravity corrections can be represented as effective modifications of nonrelativistic
Hamiltonians, without requiring commitment to a specific microscopic completion
\cite{DasVagenas2008}. This universality result provides the central motivation for applying
GUP-inspired deformations to the polaron--molecule problem: if leading quantum-gravity corrections
can be encoded at the level of effective nonrelativistic dynamics, then a many-body impurity system
described by an effective Hamiltonian offers a natural arena in which to test robustness and
sensitivity to such deformations. In that perspective, Planck-scale effects appear as higher-order
momentum-dependent contributions to the effective dynamics, allowing one to analyze low-energy
consequences of MDR/GUP-like deformations in a model-independent way. This viewpoint is
particularly natural when the system of interest is itself described by an effective Hamiltonian,
since in that case UV modifications should be introduced at the level of the effective theory rather
than through a fundamental deformation of the underlying operator algebra.

A paradigmatic example of sensitivity-driven phenomenology is provided by the trans-Planckian
problem in black-hole physics. In seminal works, ultraviolet-modified dispersion relations were
introduced phenomenologically to test the robustness of Hawking radiation against unknown
high-frequency physics, without assuming any specific ultraviolet completion
\cite{Unruh1995,CorleyJacobson1996}. These analyses showed that low-energy observables may remain robust under broad classes of UV modifications while, at the same time, providing a
controlled framework in which UV sensitivity can be quantified. The same logic has been developed in
analogue-gravity systems, where relativistic infrared behavior coexists with nonrelativistic
high-energy dispersion and the role of short-distance structure becomes both explicit and
experimentally accessible
\cite{BarceloLiberatiVisser2011}.

More generally, quantum-gravity phenomenology has increasingly adopted the view that MDR and GUP are
best regarded not as unique predictions of a single underlying theory, but as effective
parametrizations of possible UV behavior shared by broad classes of microscopic scenarios
\cite{AmelinoCamelia2013,Hossenfelder2013,LiberatiVisser2013}. From this standpoint, the central
question is not whether one particular deformation is realized in nature, but rather which physical
systems amplify sensitivity to UV modifications and which observables remain robust under changes in
the ultraviolet completion.

In parallel, low-energy quantum platforms have emerged as promising settings for this type of
analysis. Ultracold Fermi gases provide controlled realizations of effective many-body
Hamiltonians with tunable interactions and experimentally accessible probes, while quantum computing
architectures offer complementary access to real time many-body dynamics in regimes where classical
methods may become challenging
\cite{Preskill2018,CatalaValeroRodrigo2026}. In this sense, quantum computing can be used as a
laboratory for phenomenology: not as a direct probe of quantum gravity, but as a controlled setting
in which deformed effective dynamics can be implemented and their observable consequences analyzed.

In this work, we adopt this phenomenological viewpoint in a strongly correlated many-body setting.
Specifically, we consider the polaron--molecule problem, a impurity system in resonant Fermi gases
\cite{Chevy2006,GurarieRadzihovsky2007,ChinEtAl2010,KnapEtAl2012}, and use its effective
Hamiltonian as the starting point for a sensitivity analysis of MDR/GUP-like UV deformations. The
model contains an explicit ultraviolet regularization and is therefore understood throughout as an
effective low-energy description rather than a fundamental theory.

Rather than deforming canonical commutation relations at the microscopic level, we introduce
controlled UV deformations directly in the effective Hamiltonian. These deformations are constructed
to preserve the infrared sector by design, thereby defining families of effective UV completions
that agree at low energies while differing in their short-distance structure. This strategy follows
the logic of both quantum-gravity phenomenology and analogue-gravity studies, where one probes
robustness and sensitivity to UV physics without detailed knowledge of the microscopic completion
\cite{Unruh1995,CorleyJacobson1996,BarceloLiberatiVisser2011,LiberatiVisser2013}.

Our aim is therefore quantify the sensitivity of quantum gravity effects at low energies within relevant strong correlated many-body system.

To this end, we analyze spectral and Ramsey-type observables across
the polaron--molecule crossover, and we use quantum-computing implementations of the model as a
controlled laboratory for the corresponding phenomenology
\cite{CatalaValeroRodrigo2026}. We identify regimes in which many-body correlations amplify the
response to UV deformations, making their effects, in principle, resolvable at accessible energy
scales.
\subsection{Scope and claims}

The Hamiltonian studied here is an effective low-energy description, and its ultraviolet completion
is not unique. Our analysis is therefore phenomenological and sensitivity-driven. We introduce
minimal UV deformations of the effective Hamiltonian that are compatible with the same infrared
physics and examine how measurable observables respond to them. The question addressed is one of
resolvability: whether MDR/GUP-like UV effects, if present at some effective scale, could leave
detectable imprints in impurity spectroscopy and Ramsey interferometry, and whether quantum
simulation on NISQ-era devices can provide a controlled platform for exploring such effects.

The central claims of this work are the following. First, controlled UV deformations of an
effective impurity Hamiltonian can be defined consistently without altering the infrared sector.
Second, near the polaron--molecule crossover, the observables display enhanced sensitivity to
such deformations. Third, this sensitivity-based perspective, supplemented by quantum-computing
implementations as a laboratory for phenomenology, provides a conceptually consistent route for
probing quantum-gravity-inspired UV effects in low-energy many-body systems.
\section{Base effective model and observables}

To phenomenologically investigate quantum gravity signatures at low energies using  NISQ devices~\cite{Preskill2018, CatalaValeroRodrigo2026}, it is imperative to establish a rigorous theoretical bridge between a continuous field theory deformed by the  GUP and a discrete lattice model. The potency of this approach lies in the fact that the quantum circuit topology remains unaltered; conversely, the kinematic and dynamical effects of the Planck scale are encoded directly as systematic modifications to the lattice parameters and the rotation angles of the logical gates~\cite{CatalaValeroRodrigo2026}.

\subsection{Effective Hamiltonian for the two-channel model}
\label{sec:hamiltonian}
We start with a two-channel model describing a two-component fermion gas (spin $\uparrow, \downarrow$) of mass $m$, coupled to a closed molecular bosonic field $\hat{b}$ representing an impurity of mass $M=2m$~\cite{GurarieRadzihovsky2007, ChinEtAl2010}. To incorporate the GUP phenomenology, we build upon the modified algebra~\cite{KempfManganoMann1995}, where the canonical commutation relation is generalised to $[\hat{x}_i, \hat{p}_j] = i(\delta_{ij} + \beta \hat{p}^2 \delta_{ij} + 2\beta \hat{p}_i \hat{p}_j)$. This algebra, which postulates a minimum measurable length, induces modifications to the kinetic operators by introducing fourth-order spatial derivatives~\cite{DasVagenas2008}.

The total unified Hamiltonian in the continuous limit takes the form $\hat{H} = \hat{H}_\psi + \hat{H}_b + \hat{H}_{int}$. Working in natural units ($\hbar = 1$), the fermionic sector is defined as:
\begin{equation}
    \hat{H}_{\psi} = \int d^3 x \sum_{\sigma}\hat{\psi}^{\dagger}_\sigma(x) \left(-\frac{\nabla^2}{M} + \frac{\beta}{m}\nabla^4 \right)\hat{\psi}_{\sigma}(x),
    \label{eq:H_psi}
\end{equation}
where $\beta$ is the phenomenological GUP deformation parameter. Assuming universal mass scaling ($M=2m$), the molecular bosonic sector experiences a corresponding kinetic correction:
\begin{equation}
    \hat{H}_b = \int d^3 x \, \hat{b}^{\dagger}(x) \left(\nu -\frac{\nabla^2}{2M} + \frac{\beta}{M}\nabla^4 \right)\hat{b}(x),
    \label{eq:H_b}
\end{equation}
with $\nu$ denoting the bare detuning energy of the molecular state~\cite{GurarieRadzihovsky2007}. Finally, the interaction couples the channels via a Feshbach resonance:
\begin{equation}
\hat{H}_{int}=g_{bf}\int d^3 x\left(\hat{b}^{\dagger}(x)\hat{\psi}_{\downarrow}(x)\hat{\psi}_{\uparrow}(x) + \text{h.c.} \right),
\label{eq:H_int}
\end{equation}
where $g_{bf}$ represents the interconversion strength. Through the adiabatic elimination of the molecular field a valid approximation for broad resonances or large detunings $\nu$ the dynamics reduce to an effective single-channel theory~\cite{CatalaValeroRodrigo2026}.

\subsection{GUP/MDR-like deformation of the effective interaction}

We now incorporate the GUP/MDR-like sector into the two-channel framework introduced in Sec.~\ref{sec:hamiltonian}. In our notation, integrating out the bare molecular field yields the frequency-dependent effective kernel:
\begin{equation}
    g_{\mathrm{eff}}(\omega)
    =
    g_{bg}
    +
    \frac{g_{bf}^2}{\omega-(\nu-2\bar\mu)+i0^+},
    \label{eq:geff_kernel_intro}
\end{equation}
where $g_{bg}$ and $g_{bf}$ represent the background and resonant coupling strengths, respectively. To provide a clear physical interpretation for the reader, these parameters are related to the background scattering length $a_{bg}$ and the resonance width $\Delta B$ via:
\begin{equation}
    g_{bg} = \frac{4\pi a_{bg}}{m}, \quad g_{bf} = \sqrt{\frac{4\pi }{m} \frac{ \gamma \Delta B}{k_{res}}},
    \label{eq:coupling_definitions}
\end{equation}
where $m$ is the reduced mass, $\gamma$ is the difference in magnetic moments between the open and closed channels, and $k_{res}$ is the characteristic resonant wavevector. 

The first term in Eq.~\eqref{eq:geff_kernel_intro} accounts for the direct, off-resonant interactions within the open channel, while the second term captures the interaction mediated by the closed molecular channel, which is shifted by the bare detuning $\nu$. In the GUP-deformed framework, the background interaction remains essentially governed by the infrared (IR) physics, while the resonant sector will be propagated through the modified pair susceptibility $\Pi_\beta(\Omega)$ to derive the ladder-dressed coupling $g_{eff}^{(\beta)}(\Omega)$ used in the simulator.
In the present construction, the GUP correction is not introduced as an independent interaction vertex. Instead, it is implemented at the level of the open-channel kinematics and then propagated through the pair susceptibility. The resulting ladder-dressed interaction will be denoted by \(g_{\mathrm{eff}}^{(\beta)}\).

In the minimal scheme adopted here, the deformation is restricted to the open channel,
\begin{equation}
    \hat H_F^{(\beta)}
    =
    \sum_{\sigma}
    \int d^3r\,
    \hat\psi_\sigma^\dagger(\mathbf r)
    \left(
        -\frac{\nabla^2}{M}
        +
        \frac{\beta}{m}\nabla^4
        -
        \mu_\sigma
    \right)
    \hat\psi_\sigma(\mathbf r),
\end{equation}
so that the single-particle dispersion becomes
\begin{equation}
    \varepsilon_\beta(\mathbf k)
    =
    \frac{k^2}{M}
    +
    \frac{\beta}{m}k^4.
    \label{eq:eps_beta_intro_consistent}
\end{equation}
The corresponding retarded propagator is
\begin{equation}
    G_\beta^R(\mathbf k,\omega)
    =
    \frac{1}{\omega-\varepsilon_\beta(\mathbf k)+i0^+}.
\end{equation}
The density still fixes \(k_F=(6\pi^2n)^{1/3}\), and therefore
\begin{equation}
    E_F^{(\beta)}
    =
    \frac{k_F^2}{M}
    +
    \frac{\beta}{m}k_F^4.
\end{equation}
The deformation thus enters the effective interaction entirely through the modified two-particle propagator and the associated pair bubble, as shown below.

For the deformed dispersion in Eq.~\eqref{eq:eps_beta_intro_consistent}, the two-particle propagator in the ladder approximation is
\begin{equation}
    \mathcal G^{(2)}_\beta(\mathbf k,\mathbf P;\Omega)
    =
    \frac{1}{
        \Omega
        -
        E_{\mathrm{kin}}^{\mathrm{GUP}}(\mathbf k,\mathbf P)
        + i0^+
    },
    \label{eq:G2beta_main_consistent}
\end{equation}
with \(\mathbf P\) the center-of-mass momentum, \(\mathbf k\) the relative momentum, and
\begin{equation}
    E_{\mathrm{kin}}^{\mathrm{GUP}}(\mathbf k,\mathbf P)
    =
    \frac{P^2}{2M}
    +
    \frac{k^2}{m}
    +
    \frac{2\beta}{m}
    \left[
        \left(
            \frac{P^2}{4}+k^2
        \right)^2
        +
        (\mathbf k\!\cdot\!\mathbf P)^2
    \right].
    \label{eq:Ekin_main_consistent}
\end{equation}
The associated pair bubble is
\begin{equation}
    \Pi_\beta(\mathbf P,\Omega)
    =
    \int\frac{d^3k}{(2\pi)^3}\,
    \mathcal G^{(2)}_\beta(\mathbf k,\mathbf P;\Omega).
    \label{eq:Pi_beta_main_consistent}
\end{equation}
The term \((\mathbf k\!\cdot\!\mathbf P)^2\) mixes relative and center-of-mass motion and therefore breaks exact Galilean invariance at finite \(\mathbf P\). In the near-threshold sector relevant here, however, the dominant contribution comes from small center-of-mass momentum, and we accordingly expand around \(\mathbf P=0\),
\begin{equation}
    \Pi_\beta(\mathbf P,\Omega)
    =
    \Pi_\beta(\mathbf 0,\Omega)
    +
    \frac{P^2}{2}
    \left.
        \frac{\partial^2\Pi_\beta}{\partial P^2}
    \right|_{P=0}
    +
    \mathcal O(P^4).
    \label{eq:PiPexp_main_consistent}
\end{equation}

The static component is
\begin{equation}
    \Pi_\beta(\Omega)
    \equiv
    \Pi_\beta(\mathbf 0,\Omega)
    =
    \Pi_0(\Omega)
    +
    \beta\,\delta\Pi(\Omega)
    +
    \mathcal O(\beta^2),
    \label{eq:Pi_beta_expand_main_consistent}
\end{equation}
with explicit expressions given in Appendix~\ref{app:GUPbubble_consistent}. The GUP-dressed effective interaction is then defined by dressing the kernel \(g_{\mathrm{eff}}(\Omega)\) of Eq.~\eqref{eq:geff_kernel_intro} with the modified pair bubble,
restricting to the small-\(\mathbf P\) sector and expanding to first order in \(\beta\): 
\begin{equation}
    g_{\mathrm{eff}}^{(\beta)}(\Omega)
    =
    g_{\mathrm{eff}}^{(0)}(\Omega)
    \left[
        1
        +
        \beta\,g_{\mathrm{eff}}^{(0)}(\Omega)\,\delta\Pi(\Omega)
    \right]
    +
    \mathcal O(\beta^2),
    \label{eq:geffbeta_expand_main_consistent}
\end{equation}
where
\begin{equation}
    g_{\mathrm{eff}}^{(0)}(\Omega)
    =
    \left[
        g_{\mathrm{eff}}^{-1}(\Omega)
        -
        \Pi_0(\Omega)
    \right]^{-1}.
    \label{eq:geff0_main_consistent}
\end{equation}
Equation~\eqref{eq:geffbeta_expand_main_consistent} is the quantity to be matched onto the effective lattice interaction.

\subsection{Ramsey interferometry and the attractive quantum gravity regime}

To reconstruct the spectral density and probe the polaron--molecule transition~\cite{Chevy2006}, we implement a Ramsey interferometry protocol using an ancilla qubit~\cite{KnapEtAl2012}. This protocol measures the temporal coherence overlap
\begin{equation}
    S(t)
    =
    \langle \Psi_{0} |
    e^{i\hat{H}_{0}t}
    e^{-i\hat{H}_{\mathrm{eff}}^{\mathrm{GUP}}t}
    | \Psi_{0} \rangle,
\end{equation}
from which the relevant spectral information can be reconstructed.

Within the present effective description, the GUP deformation modifies the interaction sector through the \(\beta\)-dependent dressing of the effective coupling. In particular, the associated form-factor suppression reduces the weight of large-momentum scattering processes and reshapes the resonance structure. Over the parameter range accessible to the simulator, this manifests itself as an effective shift toward the molecular side of the crossover, thereby enhancing pairing tendencies and the spectral weight of bound-state-like configurations. In the NISQ implementation, the controlled variation of \(\beta\) enters through the phase
\begin{equation}
    \theta_{\mathrm{GUP}}
    =
    \frac{\tilde{U}_{\mathrm{imp}}^{\mathrm{GUP}}\Delta t}{2},
\end{equation}
and produces a measurable Ramsey phase shift. In this sense, the device acts as an analog platform for probing how ultraviolet kinematical deformations can influence low-energy pairing dynamics.


\section{Quantum computing model}
To translate the continuous effective field theory into a framework suitable for NISQ devices, we must discretise the spatial continuum. This maps the fermionic and impurity degrees of freedom onto a finite lattice, embedding the UV deformations directly into the digital quantum circuit.

\subsection{The extended Hubbard model with next-nearest-neighbour hopping}
In standard tight-binding approximations, the kinetic operator $\frac{-\nabla^2}{M}$ restricts particle hopping to nearest neighbours. However, the introdcution of the GUP induces a fourth order spatial derivative, $\frac{\beta}{m}\nabla^4$. Upon discretising the continuous spatial coordinate x into a lattice of spacing b, the finite difference expansions of this higher order derivative naturally generates next-nearest-neighbour (NNN) hopping amplitudes.

Consequently, the continuous model maps onto an extended GUP-deformed Hubbard Hamiltonian:
\begin{align}
    \hat{H}_{NISQ}^{GUP} =& -t \sum_{\langle i,j\rangle,\sigma} \hat{c}_{i\sigma}^{\dagger}\hat{c}_{j\sigma} - t^{\prime}(\beta) \sum_{\langle\langle i,j\rangle\rangle,\sigma} \hat{c}_{i\sigma}^{\dagger}\hat{c}_{j\sigma} - \tilde{t}_{imp}(m^{*}) \sum_{\langle i,j\rangle} \hat{d}_{i}^{\dagger}\hat{d}_{j} \nonumber \\
    & + U_{ff} \sum_{i} \hat{n}_{i\uparrow}\hat{n}_{i\downarrow} + \tilde{U}_{imp}^{GUP} \sum_{i} \hat{n}_{imp}(\hat{n}_{i\uparrow} + \hat{n}_{i\downarrow}),
\label{eq:HamiltonianoHubbardGUP}
\end{align}
where $\hat{c}^{(\dagger)}_{i\sigma}$ and $\hat{d}^{(\dagger)}_{i}$ are the fermionic and impurity annihilation operator at site i, and $\hat{n}$ represents the corresponding number operators.
The signatures of the UV deformation manifest across three distinct physical channels within th lattice.

The signatures of the ultraviolet deformation manifest across three distinct physical channels within the lattice:
\begin{enumerate}
    \item \textbf{Next-nearest-neighbour hopping $t^{\prime}(\beta)$:} A direct kinematic consequence of the $\beta\nabla^4$ operator, enabling non-local tunnelling across the fermionic bath.
    \item \textbf{Modified impurity hopping $\tilde{t}_{imp}(m^{*})$:} The anomalous effective mass $m^*$, derived from the $\mathcal{O}(q^2)$ Taylor expansion of the polarisation bubble, renormalises the inertia of the quasiparticle on the lattice ($\tilde{t}_{imp} \propto  / M^*b^2$).     \item \textbf{Regularised lattice interaction $\tilde{U}_{imp}^{GUP}$:} Extracted from the static limit ($q\to0$) of the dressed coupling. The continuous interaction is matched to the discrete lattice parameter via the 3D regularisation constant $\mathcal{R}_{3D}$, such that $1/\tilde{U}_{imp}^{GUP} = 1/\tilde{g}_{eff}(0,\omega) - \mathcal{R}_{3D}$.
\end{enumerate}

\subsection{System parametrisation via the Jordan-Wigner transformation}

We parametrize the fermionic anti-commutation relations on a register of distinguishable qubits with the non-local Jordan-Wigner Transformation (JWT). The fermionic operators are mapped to Pauli matrices ($X, Y, Z$) as follows:
\begin{equation}
    \hat{c}_{j}^{\dagger} = \frac{1}{2}(\hat{X}_{j} - i\hat{Y}_{j}) \otimes \bigotimes_{k=0}^{j-1} \hat{Z}_{k}.
\end{equation}
The core of our phenomenological investigation lies in the interacting sector. Under the JWT, the density-density interactions decompose into strings of Pauli-Z operators. Specifically, the GUP-deformed impurity-bath interaction maps to:
\begin{equation}
    \tilde{U}_{imp}^{GUP} \hat{n}_{imp}\hat{n}_{i\sigma} = \frac{\tilde{U}_{imp}^{GUP}}{4} \left( \mathbb{I} - \hat{Z}_{imp} - \hat{Z}_{i\sigma} + \hat{Z}_{imp}\hat{Z}_{i\sigma} \right).
\end{equation}
This diagonal mapping ensures that the physical sensitivity of the interaction to Planck-scale deformations is directly parameterised by the rotation angles of $ZZ$ coupling gates in the quantum circuit.

\subsection{Circuit design, topological layout, and Ramsey interferometry}

\begin{figure*}[th]
    \centering
    \includegraphics[width=1\linewidth]{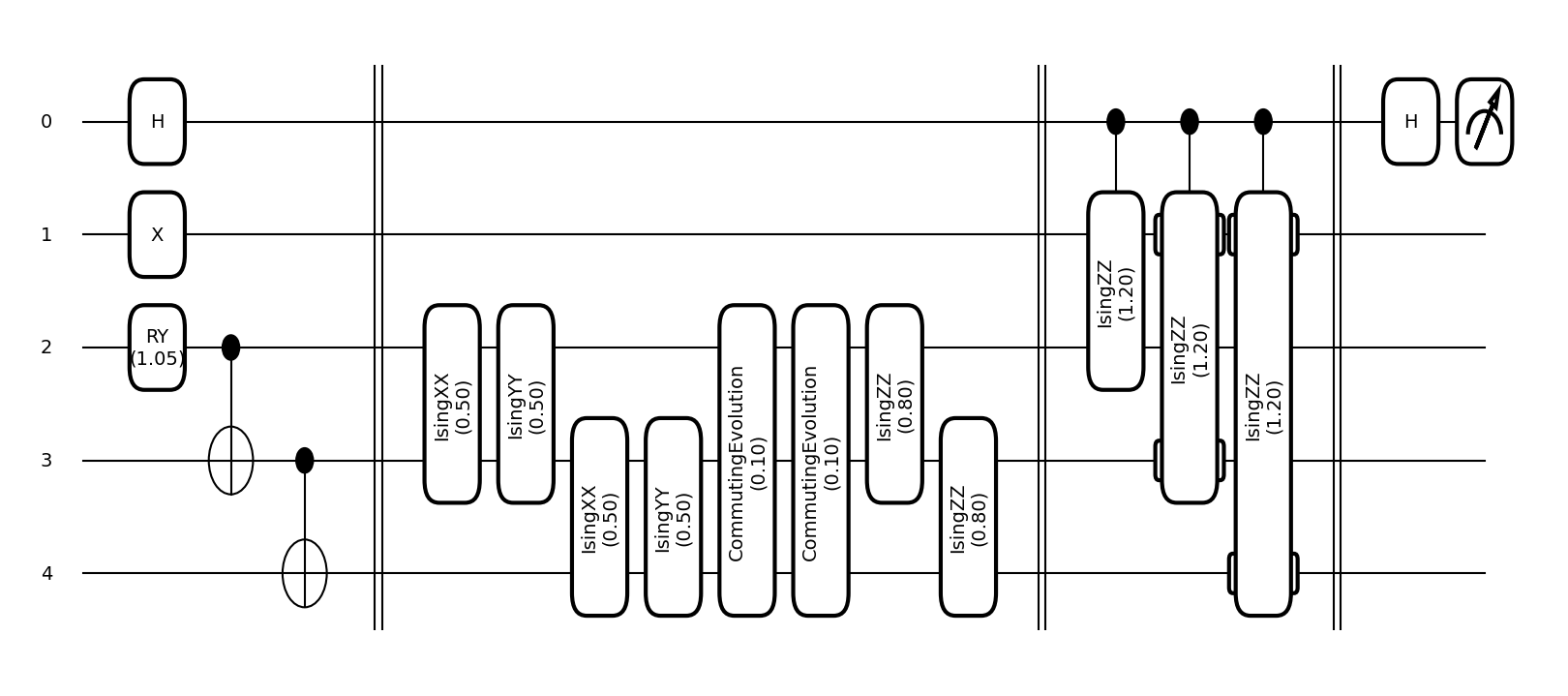}
    \caption{\textbf{Quantum circuit diagram for the ancilla-controlled Ramsey interferometry protocol.} The ancilla qubit is prepared in a superposition state via a Hadamard gate to drive a differential time evolution. The algorithmic topology is strategically partitioned: the unconditional bath evolution is driven by standard hopping, Generalised Uncertainty Principle (GUP)-induced next-nearest-neighbour hopping, and background Cooper pairing. Concurrently, the conditional impurity dynamics, which include the renormalised inertia $t_{imp}$ and the deformed interaction $\tilde{U}_{imp}^{GUP}$, are strictly controlled by the ancilla state. A final Hadamard gate enables the extraction of the real part of the coherence overlap $S(t)$ from the expectation value of the Pauli-Z operator on the ancilla.}
    \label{fig:quantikz_circuit}
\end{figure*}
The non-commuting nature of the kinetic and interacting terms ($[\hat{H}_{kin}, \hat{H}_{int}] \neq 0$) necessitates a first-order Trotter-Suzuki decomposition to approximate the unitary time evolution $\hat{U}(t) \approx (\prod e^{-i \hat{h}_k \Delta t})^{N_{steps}}$. To minimise circuit depth and mitigate gate errors, transverse correlations $\hat{Z}_{imp}\hat{Z}_{i\sigma}$ are executed via hardware-efficient MultiRZ gates, comprising a parameterised $R_Z$ rotation flanked by CNOT gates. It is at this juncture that the physics of the Planck scale is rigorously translated into machine instructions; the rotation angle of the pulses driving the entanglement is strictly defined by the GUP-renormalised interaction:
\begin{equation}
    \theta_{GUP} = \frac{\tilde{U}_{imp}^{GUP}\Delta t}{2}.
    \label{eq:theta_GUP}
\end{equation}

To probe spectral properties and observe the polaron-molecule crossover, we deploy an ancilla-controlled Ramsey interferometry protocol. As illustrated in Fig.~\ref{fig:quantikz_circuit}, the ancilla is prepared in a superposition $\frac{1}{\sqrt{2}}(|0\rangle + |1\rangle)$ via a Hadamard gate to drive a differential time evolution. The algorithmic topology is strategically partitioned:
\begin{itemize}
    \item \textbf{Unconditional Bath Evolution:} The fermionic bath undergoes continuous evolution driven by standard hopping, GUP-induced NNN hopping, and background Cooper pairing.
    \item \textbf{Conditional Impurity Dynamics:} Operations governing the impurity renormalised inertia $\tilde{t}_{imp}$ and the deformed interaction $\tilde{U}_{imp}^{GUP}$ (Eq.~\ref{eq:theta_GUP}) are strictly controlled by the ancilla state $|1\rangle$.
\end{itemize}
A final Hadamard mixing enables the extraction of the real part of the coherence overlap $S(t) = \langle\Psi_{0}|e^{i\hat{H}_{0}t}e^{-i\hat{H}_{NISQ}^{GUP}t}|\Psi_{0}\rangle$ from the expectation value of the Pauli-Z operator on the ancilla, $\langle \hat{Z}_{ancilla} \rangle = P(|0\rangle) - P(|1\rangle)$.

The reliability of this protocol has been benchmarked on superconducting hardware, such as the BSC-CNS QRed processor. Despite NISQ constraints like finite relaxation ($T_1$) and dephasing ($T_2$) times, hardware artefacts are suppressed through Readout Error Mitigation and Zero-Noise Extrapolation (ZNE). This ensures that the measured decay in coherence is genuinely attributable to many-body physics rather than gate infidelity.

\subsection{Scalability and many-body convergence}

A fundamental prerequisite for sensitivity-driven quantum gravity phenomenology is the scalability of the digital simulation towards the thermodynamic limit. The quantum circuit design developed for this framework boasts a highly efficient linear depth scaling, $\mathcal{O}(N_{T} \times N_{sites})$, with respect to both the number of Trotter steps and the spatial lattice size~\cite{CatalaValeroRodrigo2026}. The absence of fully connected all-to-all topological requirements ensures that the algorithmic compilation remains highly amenable to the constrained connectivity of near-term superconducting architectures.

The physical validity of the extracted ultraviolet signatures relies on the convergence of the discrete Hubbard model to a true many-body regime. As the qubit register is expanded to simulate larger fermionic baths, the discretised model begins to capture critical correlation effects that fundamentally transcend the capabilities of standard mean-field ansätze~\cite{Chevy2006}. The transition from a restricted few-body system to an extended bath manifests physically as an accelerated, non-Markovian decay in the initial Ramsey coherence signal. 

This macroscopic decoherence is a direct numerical and algorithmic signature of the Anderson Orthogonality Catastrophe (AOC)~\cite{KnapEtAl2012}. Within the GUP framework, the injection of the impurity not only scatters the local fermions but collectively perturbs a macroscopically extensive number of degrees of freedom across the entire deformed lattice. Furthermore, the finite-size periodic revivals of coherence typically observed in small, isolated quantum registers are exponentially suppressed as the system scales, marking a definitive convergence towards a continuous thermodynamic spectrum. 

This scalable architecture guarantees that the UV-deformed spectral features extracted via Fourier transform from the steady-state Ramsey signal are not mere finite-size algorithmic artefacts, but robust, physical manifestations of the interplay between many-body interactions and Planck-scale geometric deformations.

\section{Results}
In this section, we present the results of the quantum simulation, analyzing the dynamics of the heavy polaron and its spectroscopy across the interaction transition. The numerical results are qualitatively compared with theoretical predictions from the functional determinant formalism~\cite{knap2012time}.

 The numerical and architectural performance of the molecular polaron model was evaluated under the influence of the Generalised Uncertainty Principle (GUP) using a 10 qubit configuration. The primary metric of interest is the Ramsey coherence signal, which serves as a proxy for the stability of the impurity within the many-body bath.

\subsection{GUP/MDR-like contribution to Benchmarking and Ramsey spectroscopy}

To verify the reliability of the digital quantum simulation and the contribution of GUP to this analyse, we perform a dual-stage validation. First, we establish a theoretical baseline for a minimal system of $N=4$ qubits using Exact Diagonalisation~(ED) of the Hamiltonian in a classical environment. This provides the difference between the exact theory and the GUP/MDR-like deformation curve by computing the matrix exponential $e^{-i\hat{H}t}$ within the full $2^4$-dimensional Hilbert space, involving two fermions in the bath, one impurity and one ancilla.
 illustrates the temporal evolution of the real part of the Ramsey signal, $\text{Re}[S(t)]$, for an interaction strength of $U_{\text{imp}}/t_{ij}=2.5$ under ideal conditions (see Table~\ref{tab:beta_results}). In this stage, the \texttt{default.qubit} simulator in \textit{PennyLane} is used in statevector mode, which bypasses measurement noise to isolate algorithmic performance.
\begin{figure}[!htbp]
    \centering
    \includegraphics[width=0.9\linewidth]{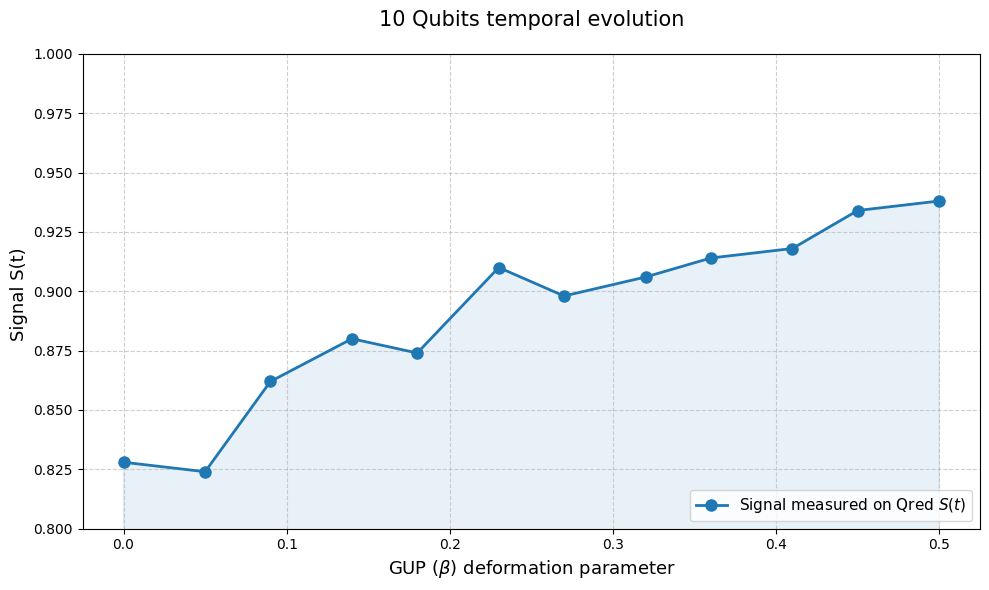}
    \caption{\textbf{Ramsey coherence signal $S(t) = \langle \hat{Z}_0 \rangle$ in a 10-qubit system as a function of the GUP deformation parameter $\beta$.} The observed increase in phase-space deformation correlates with enhanced signal preservation, suggesting that quantum gravity corrections act as a stabilisation mechanism for the polaron-molecule Hamiltonian, mitigating bath-induced decoherence, ejecuted on QRed.}
    \label{fig:gup_evolution}
\end{figure}

The evolution of the coherence signal for $\beta$ values ranging from $0.00$ to $0.50$ is documented in Table \ref{tab:beta_results}. The data reveal a clear monotonic trend, albeit with minor statistical fluctuations inherent to the quantum measurement process $N = 1000$ shots.

\begin{table}[htbp]
\centering
\caption{Measured Ramsey signal $S(t)$ values for different GUP deformation parameters $\beta$ in the 10-qubit regime.}
\label{tab:beta_results}
\begin{tabular}{cc}
\toprule
GUP parameter ($\beta$) & Signal $S(t) = \langle \hat{Z}_0 \rangle$ \\
\midrule
0.00 & 0.828 \\
0.05 & 0.824 \\
0.09 & 0.862 \\
0.14 & 0.880 \\
0.18 & 0.874 \\
0.23 & 0.910 \\
0.27 & 0.898 \\
0.32 & 0.906 \\
0.36 & 0.914 \\
0.41 & 0.918 \\
0.45 & 0.934 \\
0.50 & 0.938 \\
\bottomrule
\label{tab:Tab1}
\end{tabular}
\end{table}
Futhermore, the behaviour shown in Fig.~\ref{fig:gup_evolution} indicates that the GUP/MDR deformation modifies the Ramsey response in a physically transparent and dynamically non-trivial way. In the undeformed case, \(\beta=0\), the impurity remains more efficiently coupled to the molecular bath, so that coherence is transferred more rapidly to the many-body environment and the Ramsey signal is correspondingly more suppressed. Once the deformation is introduced, however, the time-domain response exhibits a clear increase in contrast, together with a noticeable reshaping of the oscillation pattern. Most significantly, the appearance of an additional node shows that the interferometric evolution is no longer controlled by a single characteristic frequency scale. Instead, the signal reflects a richer superposition of dynamical phases, signalling that the GUP correction changes not only the magnitude of the response but also the structure of the coherent propagation itself.

This behaviour can be understood in terms of the GUP-induced modifications of the effective polaron-molecule Hamiltonian. The higher-derivative contribution \(\beta \nabla^4\) generates a next-nearest-neighbour hopping term \(t'(\beta)\), which opens non-local propagation channels that are absent in the standard theory. At the same time, the anomalous effective mass \(m^*\), obtained from the \(\mathcal{O}(q^2)\) expansion of the polarisation bubble, renormalises the impurity hopping \(\tilde{t}_{\mathrm{imp}}(m^*)\) and therefore changes the timescale for phase accumulation during the coherent evolution. In parallel, the regularised dressed interaction \(\tilde{U}_{\mathrm{imp}}^{\mathrm{GUP}}\) modifies the impurity-bath coupling and redistributes the spectral weight between the polaronic and molecular sectors. Taken together, these effects reshape the effective dispersion relation and alter the interference pattern underlying the Ramsey trace. The resulting enhancement of the maxima and minima, combined with the emergence of an extra zero in the signal, therefore provides a clear dynamical signature of GUP/MDR physics in the polaron-molecule regime.

\begin{figure*}[t]
    \centering
    \includegraphics[width=\textwidth]{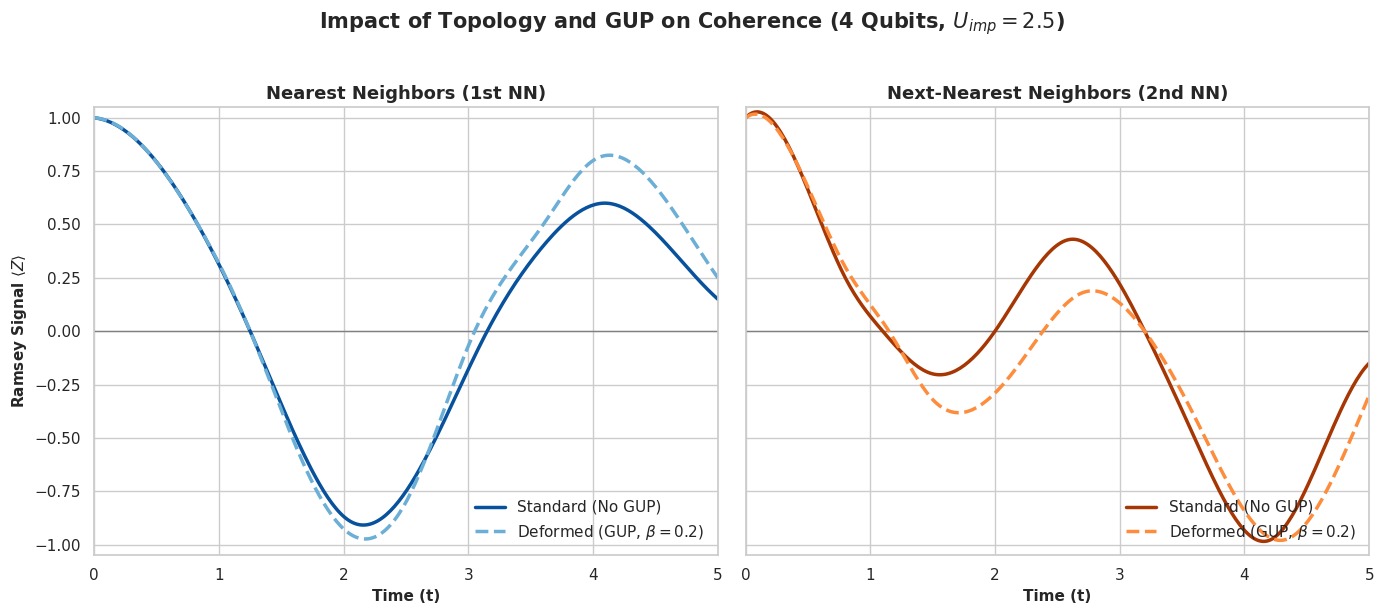}
    \caption{\textbf{Ramsey signal for the effective polaron-molecule system in the absence and presence of GUP/MDR corrections.} In the undeformed case, the stronger coupling to the molecular bath leads to a more pronounced suppression of the coherent response. Once the deformation is switched on, the time-domain signal develops a larger oscillation contrast and an additional node. This behaviour is consistent with the emergence of the GUP-induced next-nearest-neighbour hopping \(t'(\beta)\), together with the renormalisation of the effective impurity hopping \(\tilde{t}_{\mathrm{imp}}(m^*)\) and the dressed interaction \(\tilde{U}_{\mathrm{imp}}^{\mathrm{GUP}}\), all of which contribute to reshape the interferometric dynamics, executed on the QRed cluster.}
    \label{fig:gup_evolution}
\end{figure*}

\subsection{GUP/MDR-induced modifications of the quantum spectroscopy and amplitude density}
\begin{figure}
    \centering
    \includegraphics[width=1\linewidth]{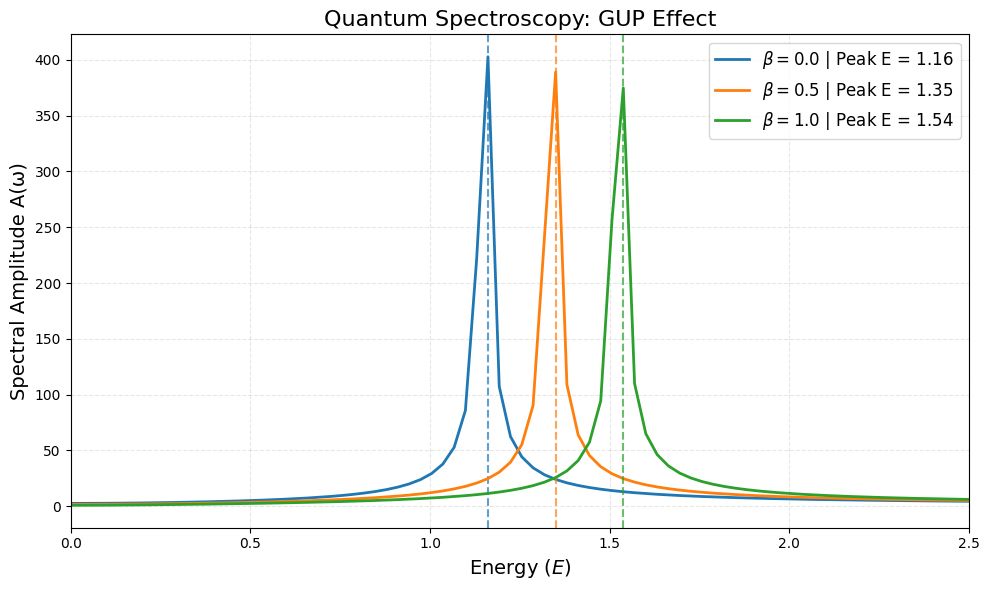}
    \caption{\textbf{Energy spectrum extracted via Fast Fourier Transform (FFT) of the time-domain Ramsey signal, illustrating the impact of the Generalised Uncertainty Principle (GUP) effect.} A systematic displacement of the resonance towards higher energies is observed as the deformation parameter $\beta$ increases. This rightward shift indicates that the effective impurity-medium sector becomes progressively stiffer, as the deformation renormalises the local dressed interaction rather than simply shifting a bare parameter. Furthermore, the concurrent reduction in peak amplitude demonstrates that the spectral weight becomes less concentrated around the resonant configuration, acting as an energetic penalty on short-range, high-momentum scattering processes.}
    \label{fig:energyfft}
\end{figure}
By applying a Fast Fourier Transform (FFT)~\cite{stewart2008using} to the Ramsey signal in Fig. \ref{fig:gup_evolution}, we extract the energy spectrum shown in Fig~\ref{fig:energyfft}. This procedure is analogous to Radio-Frequency (RF) spectroscopy used in ultracold gas experiments~\cite{chin2010feshbach, schirotzek2009observation}. To accurately localize the peak centers from the short-time Ramsey evolution, we applied Zero-Padding to the time-domain signal prior to the Fast Fourier Transform, artificially interpolating the spectral density grid without altering physical frequencies ~\cite{CatalaValeroRodrigo2026}. The resulting spectral peaks~\cite{bruun2010spectral} allow us to identify the transition from the polaron regime to the molecular state as the interaction $U_{imp}$ is tuned~\cite{Wang_2022, Mizukami2023}.

The behavior shown in Fig.~\ref{fig:energyfft}. follows directly from the matching condition
\[
\frac{1}{\tilde U_{\mathrm{imp}}^{\mathrm{GUP}}}
=
\frac{1}{\tilde g_{\mathrm{eff}}(0,\omega)}-R_{3\mathrm{D}},
\]
which maps the continuum interaction onto the discrete lattice description through the three-dimensional regularization constant \(R_{3\mathrm{D}}\). Since \(R_{3\mathrm{D}}\) removes the universal ultraviolet contribution, the full \(\beta\)-dependence of the renormalized impurity coupling is encoded in the effective vertex \(\tilde g_{\mathrm{eff}}(0,\omega)\). Consequently, increasing \(\beta\) does not simply shift a bare parameter, but rather renormalizes the local dressed interaction itself. In the spectral response, this is reflected in a systematic displacement of the resonance toward higher energies, as observed in the figure, indicating that the effective impurity-medium sector becomes progressively stiffer as the GUP deformation increases. At the same time, the reduction of the peak height shows that the spectral weight becomes less concentrated around the resonant configuration, consistent with a weaker or less sharply localized effective coupling. In this sense, the figure provides direct evidence that the GUP-induced modification of \(\tilde g_{\mathrm{eff}}(0,\omega)\), once regularized by \(R_{3\mathrm{D}}\), acts as an additional energetic penalty on short-range, high-momentum processes, thereby shifting the effective eigenvalue and attenuating the maximum spectral response.

\begin{figure*}[t!]
    \centering
    \includegraphics[width=1\linewidth]{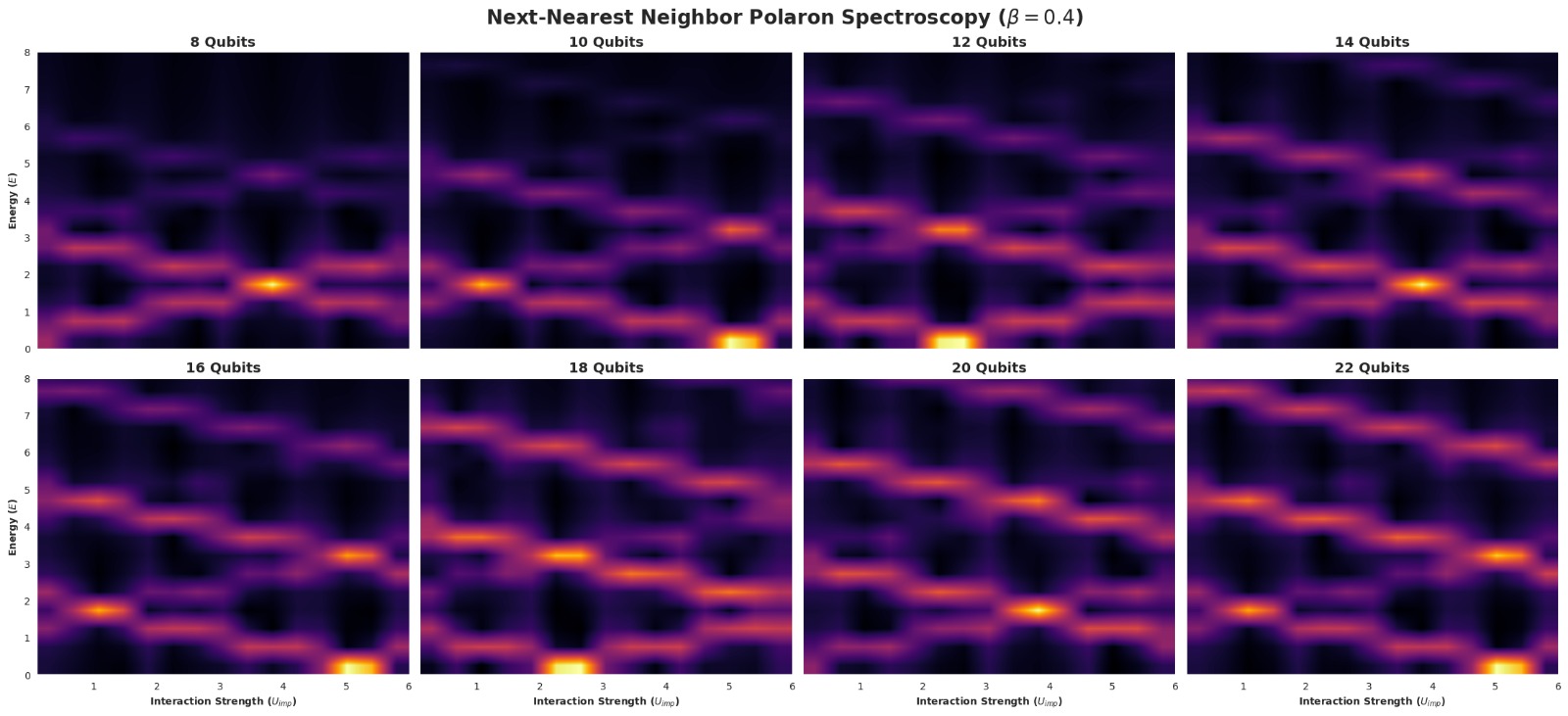}
    \caption{\textbf{Spectroscopic phase diagram.} The horizontal axis represents the interaction strength $U_{imp}$ and the vertical axis denotes the polaron energy $E_{\text{pol}}$, both expressed in units of the hopping amplitude $t_{ij}$. The linear energy shift at high coupling (bright upper branch) signals the transition to the molecular bound state. Executed on the QRed cluster.}
    \label{fig:heatmap}
\end{figure*}

In addition, is the spectroscopic phase diagram shown in Fig.~\ref{fig:branch} and Fig.~\ref{fig:heatmap}, by sweeping the effective impurity--bath interaction \(U_{\mathrm{imp}}(\beta)/t_{ij}\) over the interval \([0.1,5.0]\), where \(U_{\mathrm{imp}}\) is no longer an independent bare parameter but is controlled by the deformation parameter \(\beta\) of the underlying MDR/GUP-like framework. In this generalized setting, increasing \(\beta\) renormalizes the impurity coupling and modifies the kinetic structure of the bath, thereby shifting the onset of bound-state formation and reshaping the excitation spectrum. The heatmap reveals the continuous evolution of the many-body ground-state response across distinct spectroscopic regimes:
\begin{figure*}[!t]
    \centering
    \includegraphics[width=1\linewidth]{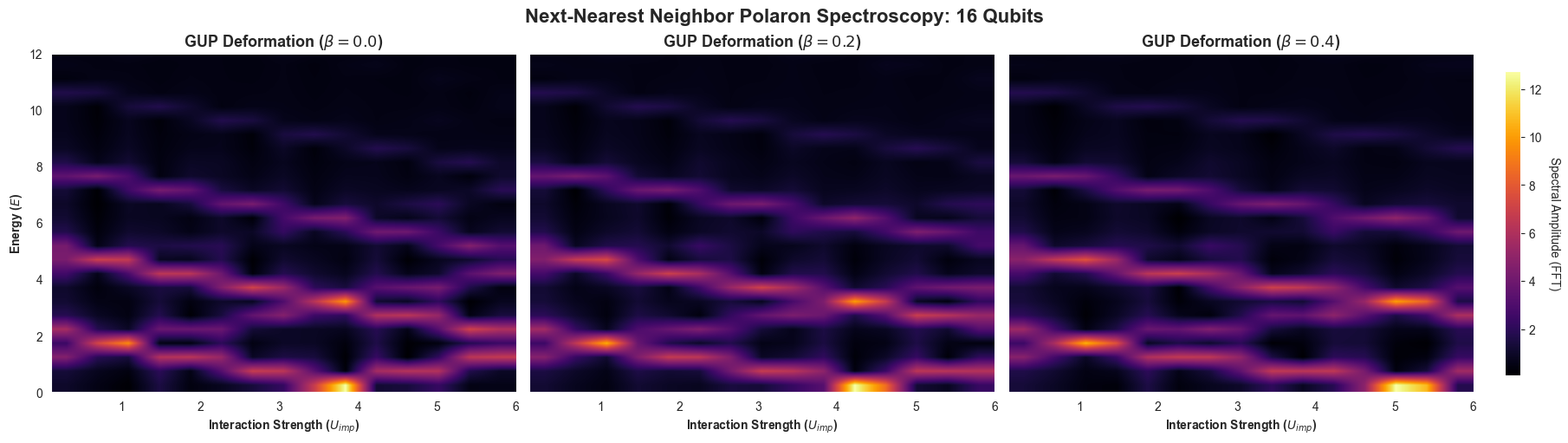}
   \caption{\textbf{Contribution of MDR/GUP-like physics in the spectroscopic phase diagram.} The horizontal axis represents the interaction strength $U_{imp}$ and the vertical axis denotes the polaron energy $E_{\text{pol}}$, both expressed in units of the hopping amplitude $t_{ij}$. The linear energy shift at high coupling signals the transition to the molecular bound state. Executed on the QRed cluster.}
    \label{fig:branch}
\end{figure*}
\begin{enumerate}
    \item \textbf{Polaron regime (\(U_{\mathrm{imp}}(\beta) \ll U_{ff}\)).}
    For weak effective coupling, the impurity energy remains close to the non-interacting limit, indicating that the impurity is only weakly dressed by particle-hole excitations of the surrounding fermionic medium and retains its quasifree character~\cite{massignan2014polarons, schirotzek2009observation}. In the deformed scenario, this regime extends over a broader interaction window as \(\beta\) increases, implying that a progressively larger effective attraction is required to destabilize the polaron in favour of a bound composite state.

    \item \textbf{Molecular regime (\(U_{\mathrm{imp}}(\beta) \gg U_{ff}\)).} 
    Upon increasing \(U_{\mathrm{imp}}(\beta)\), the spectrum develops a pronounced high-intensity branch with an approximately linear energy shift, \(E_{\mathrm{pol}} \propto U_{\mathrm{imp}}(\beta)\), which constitutes the characteristic spectroscopic fingerprint of molecule formation. In this regime, the attraction becomes strong enough to bind the impurity to a bath fermion into a composite bosonic dimer~\cite{nozieres1985bose}. Importantly, as \(\beta\) increases, the threshold for dimer formation is displaced towards larger effective interaction strengths, while the molecular feature itself emerges at higher energies, signalling that the deformation inhibits binding at low coupling and renormalizes the bound-state dispersion.
\end{enumerate}

The colour scale represents the spectral amplitude \(A(\omega)\)~\cite{schirotzek2009observation}, namely the probability density for creating a single-particle excitation at energy \(\omega\), formally related to the imaginary part of the retarded Green's function and obtained operationally from the Fourier transform of the time-domain overlap function \(S(t)\). Bright regions therefore identify long-lived, well-defined quasiparticle excitations. In the standard case, for instance around \(U_{\mathrm{imp}}/t_{ij}=4\), the dominant spectral peak is associated with the formation of a ``Molecule'' or ``Dimer''~\cite{regal2004observation}, behaving as a single composite bound state~\cite{nozieres1985bose}. In the present deformed framework, however, the modification of the underlying dynamics in particular through the inclusion of next-nearest-neighbour processes generates additional high-energy spectral branches beyond the conventional molecular one. These emergent branches also display an approximately linear shift at strong coupling, indicating the appearance of new bound or resonant composite excitations induced by the interplay between interaction renormalization and altered hopping structure.

The visual separation between the bright upper branches and the lower diffuse continuum is the direct manifestation of binding: the formation of a molecular or molecule-like excitation opens an energy gap relative to the scattering continuum~\cite{nozieres1985bose}. Qualitatively, the resulting diagram reproduces the essential physics of the polaron-to-molecule transition established in the theoretical literature~\cite{massignan2014polarons} and observed experimentally~\cite{schirotzek2009observation, Mizukami2023}, while simultaneously revealing that MDR/GUP-induced corrections shift the transition point, push dimer formation to higher energies, and promote additional spectroscopic branches once longer-range dynamical processes are taken into account.

Figure~\ref{fig:heatmap} examines the stability of the deformed spectral response as the number of qubits is increased, keeping \(\beta=0.4\) fixed. The spectra retain a multibranch structure for larger registers, although the relative intensity and sharpness of individual branches vary with system size. This indicates that the additional structures generated by the next-nearest-neighbour deformation are not solely artifacts of the smallest simulations, but also that they should not all be identified with dimer formation. Several bands correspond to dressed impurity excitations, scattering-like states, or finite-size levels, whereas only those high-intensity branches that separate from the diffuse continuum and persist as the system size increases can be interpreted as molecular or bound-state-like features. In this sense, the figure supports a picture in which the MDR/GUP correction redistributes the spectral weight across multiple excitation channels, enriching the crossover structure without implying the formation of multiple independent dimers.

\subsection{Digital quantum validation}
The digital quantum validation protocol implemented in this work adheres strictly to the benchmarking and error-mitigation framework thoroughly established in our previous work~\cite{CatalaValeroRodrigo2026}. Specifically, the core algorithmic architecture, including the deployment of Readout Error Mitigation matrices and the Zero-Noise Extrapolation (ZNE) protocol used to suppress hardware artefacts, remains identical to the procedures validated therein. 

The primary departure from the baseline validation described lies in the modification of the digital circuit topology to accommodate the GUP-induced ultraviolet deformations. Instead of evaluating a standard, undeformed single-channel model, the current validation tracks how the systematically modified lattice parameters alter the machine-instruction mapping. This operational divergence is explicitly quantified in Tab.~\ref{tab:Tab1}, which documents the evolution of the Ramsey coherence signal $S(t)$ as a function of the non-zero deformation parameter $\beta$ a physical degree of freedom completely absent in our first paper. Furthermore, while the initial demonstration focused on the ideal statevector baseline and a highly restricted register, the current hardware validation utilizes an expanded 10-qubit register on the QRed superconducting processor.  

The hardware characterisation parameters reported in Tab.~\ref{tab:tabla decoherencia} reflect the updated cross-talk, gate fidelities, and decoherence times ($T_1, T_2$) corresponding to the scaled topological layout required to execute the multi-RZ gate sequences under next-nearest-neighbour hopping.

\subsection{Hardware Characterization and Reproducibility}
To ensure the reproducibility of this demonstration, we report the operational parameters of the BSC-CNS QRed quantum processor at the time of execution. The system consists of superconducting transmon qubits. Tab.~\ref{tab:tabla decoherencia} summarizes the coherence times ($T_1, T_2$), resonance frequencies, and error rates for the qubits utilized in the Ramsey protocol (Qubits $Q_0$ to $Q_20$).
\begin{table*}[t]
\centering
\caption{Coherence times ($T_1$ and $T_2$), single-qubit gate fidelities, readout fidelities, and two-qubit gate fidelities for qubits 5, 10, 15, and 20. QRed Cluster}
\label{tab:qubit_metrics}
\begin{tabular}{cccccc}
\toprule
\textbf{Qubit} & \textbf{$T_1$ ($\mu$s)} & \textbf{$T_2$ ($\mu$s)} & \textbf{1Q Gate} & \textbf{Readout} & \textbf{2Q Gate} \\
\midrule
5  & 130.79 & 110.75 & 99.90 & 93.00 & 5\_4: 87.33 - 22\_5: 89.59 \\
10 & 150.72 & 15.47 & 99.64 & 97.26 & 7\_10: 88.07 \\
15 & 120.34  & 110.47  & 99.70 & 93.44 & 13\_15: 86.01 - 15\_16: 89.23 - 18\_15: 88.37 \\
20 & 150.35 & 140.13 & 99.93 & 95.23 & 20\_19: 89.43 \\
\bottomrule
\label{tab:tabla decoherencia}
\end{tabular}
\end{table*}

The gate fidelities and coherence times reported were obtained via standard dispersive readout and randomized benchmarking protocols performed by the BSC-CNS facility. These parameters were used to calibrate the Readout Error Mitigation matrix $M$ and to set the scale for ZNE. The stability of these metrics during the 1000 shot acquisition window ensures that the observed spectral bifurcation is a physical property of the effective Hamiltonian and not a drift-induced artifact.

\section{Beyond-Mean-Field Correlations and Scalability in the Polaron--Molecule Crossover}
\label{sec:correlations_and_scalability}

The BEC--BCS transition is governed by the population imbalance and the dimensionless coupling parameter $1/(k_F a)$~\cite{giorgini2008theory}, which in our lattice simulation maps to the effective onsite interaction strengths $U_{ff}$ and $U_{imp}$. In the extreme imbalanced limit, where a single impurity interacts with a majority Fermi sea, the system forms a polaron~\cite{schirotzek2009observation}. In the weakly attractive BCS regime where bath fermions form large, overlapping Cooper pairs~\cite{eagles1969possible} the impurity is ``dressed'' by particle-hole excitations, forming a quasiparticle with a renormalised mass and energy~\cite{massignan2014polarons}. As the interaction $U_{imp}/t_{ij}$ is swept towards the strongly attractive BEC limit, $U_{imp}$ exceeds the bath's self-interaction $U_{ff}$. This drives a smooth crossover where the quasiparticle and a bath fermion shrink into a tightly bound molecular dimer~\cite{nozieres1985bose, regal2004observation, Wang_2022, parish2011polaron}.

Standard mean-field theories and conventional Chevy-like variational Ansätze~\cite{massignan2014polarons, chevy2006universal} often struggle to capture the strongly correlated nature of this crossover, particularly near the unitary limit where the scattering length diverges~\cite{bloch2008many, randeria2014crossover, zwerger2012bcs}. Our digital quantum simulation overcomes these limitations by mapping the full many-body wave function directly onto the quantum processor's Hilbert space. By evolving the system under the full Hubbard-discretised Hamiltonian, we explicitly incorporate the multi-particle scattering processes essential for accurately determining the polaron lifetime and the spectral weight transfer to the molecular branch~\cite{Wang_2022, Mizukami2023}.

A fundamental benchmark for this approach is its performance as the system size scales towards the thermodynamic limit~\cite{bloch2008many}. The introduction of an impurity typically causes a macroscopic rearrangement of the Fermi sea. By scaling the fermionic bath from $N=6$ to $N=10$ qubits, we observe a markedly steeper initial decay in the Ramsey coherence signal, providing direct numerical evidence of the \textit{Anderson Orthogonality Catastrophe}~\cite{anderson1967infrared}. The injected impurity scatters against an increasingly extensive number of lattice degrees of freedom, exponentially suppressing the overlap between the initial and time-evolved many-body states. Concurrently, while smaller restricted registers ($N=4$ and $N=6$) inherently display periodic coherence revivals due to their highly discrete energy spectra, the $N=10$ simulation heavily damps these finite-size recurrences. This suppression confirms our digital compilation approaches the irreversible dephasing characteristic of a thermodynamic continuum, even as the superfluid gap $\Delta$ acts as a protective shield that partially suppresses low-energy particle-hole excitations to stabilise polaron coherence~\cite{zwierlein2005vortices, leggett2006quantum}.

To transparently assess this scalability on current NISQ devices, it is crucial to quantify the algorithmic resource overhead. For the $N=10$ system ($M=4$ Cooper pairs), transpiling the controlled Jordan-Wigner Hamiltonian now extended with the GUP-induced next-nearest-neighbour hopping to the BSC-CNS hardware requires approximately 150 CNOT gates per Trotter step. Consequently, the optimal 15-step evolution demands a total circuit depth of $\sim 2250$ CNOTs. With an average gate duration of 150 ns, the total execution time ($\sim 337.5$ $\mu$s) clearly exceeds the average $T_2$ coherence envelope of the processor ($\sim 14$ $\mu$s). 

Whilst our ZNE protocol effectively recovers signal amplitude during the shallower initial Trotter steps successfully capturing the onset of the many-body dephasing the signal naturally converges to the device's noise floor as the execution time heavily surpasses the $T_2$ limit. This explicitly delineates the current hardware frontier, marking the exact point where state-of-the-art processors transition from successfully simulating coherent strongly-correlated dynamics to being bounded by intrinsic decoherence. However, the linear $\mathcal{O}(N)$ scaling of the circuit depth ensures that our protocol remains viable for near-term NISQ architectures. To maintain high fidelity in even larger systems, future implementations could utilise second-order Trotter-Suzuki formulae~\cite{berry2007efficient} or variational time-evolution principles~\cite{yuan2019theory} to further mitigate hardware decoherence and resolve the finer details of the polaron molecule crossover.

\section{Conclusions}
In this work, we have extended the digital quantum-simulation framework previously developed for strongly correlated fermionic impurity systems by incorporating an effective kinematic correction inspired by generalized-uncertainty-principle and modified-dispersion-relation scenarios. This GUP/MDR-like contribution deforms the single-particle kinetic sector and provides a controlled route to assess how departures from the standard low-energy dispersion affect the many-body physics of the Fermi polaron and its crossover towards molecular bound-state formation.

A central outcome of this analysis is that the effective deformation does not simply renormalize the energy spectrum. Instead, it modifies the structure of the simulated Hamiltonian by generating next-nearest-neighbour contributions, thereby opening additional dynamical pathways in the impurity-bath evolution. These longer-range terms alter the real-time dynamics of the correlated system and enrich the mechanism underlying the polaron-to-molecule transition beyond the standard nearest-neighbour description.

Using a Ramsey-interferometry protocol as a quantum-computational probe, we have shown that the deformed dynamics leaves clear signatures in experimentally motivated observables. In particular, the Ramsey coherence signal exhibits a systematic phase shift towards longer times as the deformation parameter $\beta$ is increased. This delay indicates that the GUP/MDR-like kinetic correction modifies the characteristic timescales associated with impurity dressing, coherence decay, and molecular formation. Therefore, the Ramsey response provides a sensitive diagnostic of the interplay between modified kinematics and strongly correlated many-body dynamics.

The spectral properties of the effective Hamiltonian further support this interpretation. We find that the eigenvalues increase monotonically with the GUP/MDR control parameter $\beta$, revealing an enhancement of the relevant energy scales induced by the modified dispersion relation. This $\beta$-dependent spectral restructuring confirms that the deformation acts as a genuine physical control parameter rather than as a purely perturbative correction. Consequently, the location and structure of the polaron molecule crossover are quantitatively modified.

Consistently, the probability maps associated with the polaron-molecule transition show that the regions of largest transition probability are displaced towards stronger interaction potentials as $\beta$ increases. This behaviour indicates that, in the deformed kinematic regime, larger attractive interactions are required to favour molecular bound-state formation. Moreover, the emergence of additional channels generated by next-nearest-neighbour processes modifies the transition landscape, giving rise to a richer structure than in the undeformed model. The combined observation of delayed Ramsey dynamics, spectral enhancement, displacement of high-probability transition regions, and the appearance of new dynamical channels provides a coherent physical picture of how GUP/MDR-like effects reshape the strongly correlated impurity problem.

These results establish the Fermi-polaron platform as a sensitive many-body probe of effective high-energy-inspired corrections to quantum dynamics. More broadly, they demonstrate that digital quantum simulation can be used not only to reproduce strongly correlated matter phenomena, but also to explore controlled phenomenological extensions of standard quantum many-body models. In this sense, the Ramsey response of the polaron-to-molecule transition constitutes a useful quantum-computational laboratory for investigating the observable consequences of generalized kinematic structures.

Although current quantum-hardware limitations, including decoherence, finite circuit depth, and imperfect gate fidelities, still constrain the achievable spectral resolution, the robustness and internal consistency of the observed trends support the validity of the proposed framework. The present study therefore opens a pathway towards employing gate-based quantum devices as laboratories for the investigation of strongly correlated systems subject to effective quantum gravity inspired kinematic deformations, bridging concepts from ultracold-atom physics, quantum information processing, and beyond-standard quantum dynamics. \\

\section*{Acknowledgments}

The authors thankfully acknowledge the Spanish Supercomputing Network (RES) resources provided by BSC-CNS in MareNostrum5/Quantum-Blue/Quantum-Red to FI-2025-3-0043 activity. Work supported by the Spanish Government and ERDF/EU - Agencia Estatal de Investigaci\'on (MCIU/AEI/10.13039/501100011033), Grant No. PID2023-146220NB-I00, This work is also supported by the Ministry of Economic Affairs and Digital Transformation of the Spanish Government and NextGenerationEU through the Quantum Spain project, and by CSIC Interdisciplinary Thematic Platform (PTI+) on Quantum Technologies (PTI-QTEP+). 

\bibliographystyle{iopart-num}
\bibliography{refs}

\appendix
\section{GUP-modified pair bubble and ladder-dressed effective interaction}
\label{app:GUPbubble_consistent}

In this Appendix we summarize the derivation of the GUP-modified pair bubble and of the corresponding ladder-dressed effective interaction used in the main text. Throughout, \(\mathbf P\) denotes the center-of-mass momentum, \(\mathbf k\) the relative momentum, and \(\Omega\) the pair energy.

We start from the deformed single-particle dispersion
\begin{equation}
    \varepsilon_\beta(\mathbf p)
    =
    \frac{p^2}{2m}
    +
    \frac{\beta}{m}p^4.
    \label{eq:epsbeta_app_consistent}
\end{equation}
For two particles with momenta \(\mathbf p_{1,2}=\mathbf P/2\pm\mathbf k\), the kinetic energy is
\begin{equation}
    E_{\mathrm{kin}}^{\mathrm{GUP}}(\mathbf k,\mathbf P)
    =
    \varepsilon_\beta\!\left(\frac{\mathbf P}{2}+\mathbf k\right)
    +
    \varepsilon_\beta\!\left(\frac{\mathbf P}{2}-\mathbf k\right).
\end{equation}
Using
\begin{equation}
    \left(\frac{\mathbf P}{2}\pm\mathbf k\right)^2
    =
    \frac{P^2}{4}+k^2\pm\mathbf P\!\cdot\!\mathbf k,
\end{equation}
one obtains
\begin{align}
    \left(\frac{\mathbf P}{2}+\mathbf k\right)^4
    +
    \left(\frac{\mathbf P}{2}-\mathbf k\right)^4
    &=
    \left(\frac{P^2}{4}+k^2+\mathbf P\!\cdot\!\mathbf k\right)^2
    +
\end{align}
\begin{align}
     + \left(\frac{P^2}{4}+k^2-\mathbf P\!\cdot\!\mathbf k\right)^2
    \nonumber 
    &=
    2\left[
        \left(\frac{P^2}{4}+k^2\right)^2
        +
        (\mathbf P\!\cdot\!\mathbf k)^2
    \right]
\end{align}
Therefore,
\begin{equation}
    E_{\mathrm{kin}}^{\mathrm{GUP}}(\mathbf k,\mathbf P)
    =
    \frac{P^2}{4m}
    +
    \frac{k^2}{m}
    +
    \frac{2\beta}{m}
    \left[
        \left(
            \frac{P^2}{4}+k^2
        \right)^2
        +
        (\mathbf P\!\cdot\!\mathbf k)^2
    \right],
    \label{eq:Ekin_app_consistent}
\end{equation}
which reproduces Eq.~\eqref{eq:Ekin_main_consistent}. The \((\mathbf P\!\cdot\!\mathbf k)^2\) term prevents an exact separation between relative and center-of-mass motion and thus breaks Galilean invariance at finite \(\mathbf P\).

After frequency integration, the two-particle propagator becomes
\begin{equation}
    \mathcal G^{(2)}_\beta(\mathbf k,\mathbf P;\Omega)
    =
    \frac{1}{
        \Omega
        -
        E_{\mathrm{kin}}^{\mathrm{GUP}}(\mathbf k,\mathbf P)
        + i0^+},
    \label{eq:G2_app_consistent}
\end{equation}
and the corresponding pair bubble is
\begin{equation}
    \Pi_\beta(\mathbf P,\Omega)
    =
    \int\frac{d^3k}{(2\pi)^3}\,
    \mathcal G^{(2)}_\beta(\mathbf k,\mathbf P;\Omega).
    \label{eq:Pi_app_consistent}
\end{equation}
Expanding around \(\mathbf P=0\),
\begin{equation}
    \Pi_\beta(\mathbf P,\Omega)
    =
    \Pi_\beta(\mathbf 0,\Omega)
    +
    \frac{P^2}{2}
    \left.
        \frac{\partial^2\Pi_\beta}{\partial P^2}
    \right|_{P=0}
    +
    \mathcal O(P^4).
    \label{eq:Pexp_app_consistent}
\end{equation}
The static component is
\begin{equation}
    \Pi_\beta(\Omega)
    \equiv
    \Pi_\beta(\mathbf 0,\Omega)
    =
    \int\frac{d^3k}{(2\pi)^3}
    \frac{1}{
        \Omega
        -
        \dfrac{k^2}{m}
        -
        \dfrac{2\beta}{m}k^4
        + i0^+ }.
    \label{eq:Pibeta_static_app_consistent}
\end{equation}
For \(\beta k^2\ll1\), this admits the perturbative expansion
\begin{equation}
    \Pi_\beta(\Omega)
    =
    \Pi_0(\Omega)
    +
    \beta\,\delta\Pi(\Omega)
    +
    \mathcal O(\beta^2),
    \label{eq:Pibeta_expand_app_consistent}
\end{equation}
with
\begin{equation}
    \Pi_0(\Omega)
    =
    \int\frac{d^3k}{(2\pi)^3}
    \frac{1}{
        \Omega-\dfrac{k^2}{m}+i0^+},
    \label{eq:Pi0_app_consistent}
\end{equation}
and
\begin{equation}
    \delta\Pi(\Omega)
    =
    \int\frac{d^3k}{(2\pi)^3}
    \frac{\dfrac{2k^4}{m}}{
        \left(
            \Omega-\dfrac{k^2}{m}+i0^+
        \right)^2 }.
    \label{eq:dPi_app_consistent}
\end{equation}

The undeformed two-channel kernel obtained after integrating out the bare molecular field is
\begin{equation}
    g_{\mathrm{eff}}(\Omega)
    =
    g_{bg}
    +
    \frac{g_{bf}^2}{\Omega-(\nu-2\bar\mu)+i0^+}.
    \label{eq:geff_kernel_app_consistent}
\end{equation}
The ladder-dressed GUP-corrected effective interaction is then defined by
\begin{equation}
    \bigl[g_{\mathrm{eff}}^{(\beta)}(\mathbf P,\Omega)\bigr]^{-1}
    =
    \bigl[g_{\mathrm{eff}}(\Omega)\bigr]^{-1}
    -
    \Pi_\beta(\mathbf P,\Omega).
    \label{eq:geffbeta_def_app_consistent}
\end{equation}
Specializing to \(\mathbf P=0\) and expanding to first order in \(\beta\),
\begin{equation}
    \bigl[g_{\mathrm{eff}}^{(\beta)}(\Omega)\bigr]^{-1}
    =
    \bigl[g_{\mathrm{eff}}^{(0)}(\Omega)\bigr]^{-1}
    -
    \beta\,\delta\Pi(\Omega)
    +
    \mathcal O(\beta^2),
    \label{eq:geffinv_expand_app_consistent}
\end{equation}
where
\begin{equation}
    g_{\mathrm{eff}}^{(0)}(\Omega)
    =
    \left[
        g_{\mathrm{eff}}^{-1}(\Omega)
        -
        \Pi_0(\Omega)
    \right]^{-1}.
    \label{eq:geff0_app_consistent}
\end{equation}
Perturbative inversion gives
\begin{equation}
    g_{\mathrm{eff}}^{(\beta)}(\Omega)
    =
    g_{\mathrm{eff}}^{(0)}(\Omega)
    \left[
        1
        +
        \beta\,g_{\mathrm{eff}}^{(0)}(\Omega)\,\delta\Pi(\Omega)
    \right]
    +
    \mathcal O(\beta^2),
    \label{eq:geff_expand_app_consistent}
\end{equation}
which reproduces Eq.~\eqref{eq:geffbeta_expand_main_consistent} of the main text.

Finally, the same low-\(\mathbf P\) expansion determines the inertial correction to the dressed composite mode. Writing the momentum-dependent self-energy as \(\Sigma(\mathbf P)\), the effective mass \(m^*\) is defined through
\begin{equation}
    \frac{P^2}{2m^*}
    \equiv
    \Sigma(\mathbf P)-\Sigma(\mathbf 0)
    \propto
    P^2
    \left.
        \frac{\partial^2\Pi_\beta}{\partial P^2}
    \right|_{P=0},
    \label{eq:mstar_app_consistent}
\end{equation}
so that both the quartic term in Eq.~\eqref{eq:Ekin_app_consistent} and the angular structure \((\mathbf P\!\cdot\!\mathbf k)^2\) contribute to the anomalous inertial response.

\section{How to extend Hubbard model with GUP into a NNN hopping amplitude}
\label{annex:gup_second_neighbors}

In this Appendix, we provide a comprehensive derivation of the lattice Hamiltonian used to simulate Generalized Uncertainty Principle (GUP) effects. We demonstrate how higher-order spatial derivatives, which are a indicator of ultraviolet (UV) modifications to the dispersion relation, naturally manifest as non-local hopping terms when mapped onto a discrete quantum register.

\subsection*{B1. The Continuum Deformed Kinetic Sector}

We consider a system of fermions described by a field operator $\hat{\psi}_{\sigma}(\mathbf{x})$. In the presence of a minimal measurable length scale, the standard Schrödinger kinetic energy is augmented by a leading-order GUP correction. The effective kinetic Hamiltonian density, as introduced in Eq.~\ref{eq:H_psi}, is given by:
\begin{equation}
\hat{H}_{kin} = \int d^3x \sum_{\sigma} \hat{\psi}_{\sigma}^{\dagger}(\mathbf{x}) \left[ -\frac{\nabla^2}{M} + \frac{\beta}{m} \nabla^4 \right] \hat{\psi}_{\sigma}(\mathbf{x}) \tag{B1}
\end{equation}
where $\beta$ is the GUP deformation parameter. In the context of quantum simulation, we set $\hbar=1$. The operator $\nabla^4 \equiv \nabla^2(\nabla^2)$ introduces a higher-order sensitivity to the high-momentum (UV) modes of the field.

\subsection*{B2. Lattice Discretisation and Finite Difference Stencils}

To implement this dynamics on a digital quantum computer with $L$ sites, we discretise the continuous coordinate $\mathbf{x}$ into a 1D lattice (generalisable to higher dimensions) with lattice spacing $b$. The continuous field $\hat{\psi}(x_i)$ is mapped to the discrete annihilation operator $\hat{c}_i$ via the relation $\hat{c}_i \simeq \sqrt{b} \hat{\psi}(x_i)$.

The discretisation of the spatial derivatives is performed using the central finite difference method. For the second-order derivative (the standard Laplacian), the three-point stencil at site $i$ is:
\begin{equation}
\nabla^2 \hat{c}_i \approx \frac{\hat{c}_{i+1} - 2\hat{c}_i + \hat{c}_{i-1}}{b^2}. \tag{B2}
\end{equation}
To discretise the fourth-order derivative $\nabla^4$, we apply the operator (B2) to itself:
\begin{align}
\nabla^4 \hat{c}_i &\approx \frac{1}{b^2} \left[ \nabla^2 \hat{c}_{i+1} - 2\nabla^2 \hat{c}_i + \nabla^2 \hat{c}_{i-1} \right] \nonumber \\
&= \frac{1}{b^4} \left[ \left(\hat{c}_{i+2} - 2\hat{c}_{i+1} + \hat{c}_i\right) - 2\left(\hat{c}_{i+1} - 2\hat{c}_i + \hat{c}_{i-1}\right) \right. \nonumber \\ 
& \quad + \left. \left(\hat{c}_i - 2\hat{c}_{i-1} + \hat{c}_{i-2}\right) \right]. \tag{B3}
\end{align}
Collecting terms according to their spatial separation from the central site $i$, we obtain the five-point stencil for the GUP correction:
\begin{equation}
\nabla^4 \hat{c}_i \approx \frac{\hat{c}_{i+2} - 4\hat{c}_{i+1} + 6\hat{c}_i - 4\hat{c}_{i-1} + \hat{c}_{i-2}}{b^4}. \tag{B4}
\end{equation}

\subsection*{B3. Emergence of the Extended Hubbard Parameters}

Substituting the stencils (B2) and (B4) back into the Hamiltonian (B1), and focusing on the kinetic energy contribution $\hat{H}_{kin} = \sum_i \hat{c}_i^\dagger [\hat{K}] \hat{c}_i$, we find:
\begin{align}
\hat{H}_{kin} = \sum_{i,\sigma} \hat{c}_i^\dagger & \left[ -\frac{1}{2mb^2} (\hat{c}_{i+1} - 2\hat{c}_i + \hat{c}_{i-1}) \right. \nonumber \\ 
& \left. + \frac{\beta}{mb^4} (\hat{c}_{i+2} - 4\hat{c}_{i+1} + 6\hat{c}_i - 4\hat{c}_{i-1} + \hat{c}_{i-2}) \right]. \tag{B5}
\end{align}
Grouping these by hopping distance, the Hamiltonian can be written in the standard form of an extended tight-binding model:
\begin{equation}
\hat{H}_{kin} = \epsilon_0 \sum_{i,\sigma} \hat{n}_{i\sigma} - t_{eff} \sum_{\langle i,j \rangle, \sigma} (\hat{c}_{i\sigma}^\dagger \hat{c}_{j\sigma} + h.c.) - t' \sum_{\langle\langle i,j \rangle\rangle, \sigma} (\hat{c}_{i\sigma}^\dagger \hat{c}_{j\sigma} + h.c.) \tag{B6}
\end{equation}
where the effective parameters are defined as follows:
\begin{itemize}
    \item \textbf{On-site energy shift:} $\epsilon_0 = \frac{1}{mb^2} + \frac{6\beta}{mb^4}$. This term contributes to a global shift of the chemical potential and does not affect the relative dynamics of the polaron.
    \item \textbf{Effective NN Hopping ($t_{eff}$):} The GUP term modifies the standard tunnelling amplitude $t_0 = 1/2mb^2$:
    \begin{equation}
    t_{eff} = \frac{1}{2mb^2} + \frac{4\beta}{mb^4}. \tag{B7}
    \end{equation}
    \item \textbf{NNN Hopping ($t'$):} This represents the non-local tunnelling directly induced by the $\nabla^4$ UV deformation:
    \begin{equation}
    t'(\beta) = -\frac{\beta}{mb^4}. \tag{B8}
    \end{equation}
\end{itemize}

\end{document}